\documentclass[aps,preprint,superscriptaddress,showpacs]{revtex4}

\usepackage{graphicx}
\usepackage{amsmath}
\usepackage{amssymb}
\usepackage{hyperref}
\bibstyle{apsrev.bib}
\usepackage{color}

\begin{document}

\title{About the parabolic relation existing between the skewness and the kurtosis in time series of experimental data}

\author{ F. Sattin, M. Agostini, R. Cavazzana, G. Serianni, P. Scarin, N. Vianello }

\affiliation{Consorzio RFX, Associazione EURATOM-ENEA sulla fusione, Corso Stati Uniti 4,
Padova, Italy}

\begin{abstract}
In this work we investigate the origin of the parabolic relation between skewness and kurtosis
often encountered in the analysis of experimental time-series. We argue that the numerical values of 
the coefficients of the curve  may 
provide informations about the specific physics of the system studied, 
whereas the analytical curve \textit{per se} is a fairly general consequence of a few constraints
expected to hold for most systems.     
\end{abstract}


\maketitle

The study of  turbulent, disordered or chaotic   systems must often rely on the statistical analysis
of experimental time series of one or more variables representative of the state of the system observed.
This is true for those fields where controlled experiments are hardly feasible  and one must resort on passive observations alone (e.g., geophysics and environmental sciences, econophysics, ...). Even in disciplines where
controlled experiments may be performed, the understanding of the underlying 
mechanisms may not have yet been progressed enough to account for all observed patterns, and one must follow 
the other way round, disentangling pieces of information from raw time series. This is the case, e.g., 
for some problems of fluid dynamics and plasma physics in magnetic confinement devices. \\
The most informative way of expressing the information contained in a time series is by building its Probability Distribution Function (PDF), which accounts for the frequency of the measured variable of attaining a specific value.
Rather than the whole PDF, often its moments are considered: being 
averages, moments may be estimated quite reliably from relatively small amounts of data.   
Due to the limited amount of data usually available, only the first moments are generally
computable reliably. A common finding, repeatedly verified in completely different environments,
is that the numerical values of these moments are not constant, rather may differ even appreciably between
two repetitions of the experiments. What is particularly remarkable   
is that very often a well-defined correlation exists between the third and fourth--order normalized moments: 
Skewness $S$ and Kurtosis $K$ (Throughout this paper, the word ``kurtosis'' will label the normalized fourth moment, not its difference from the gaussian value 3--''excess kurtosis''): after several measurements, one gets several copies $(S,K)$, and finds that, approximately, they align along a quadratic curve:
\begin{equation}
K = A \cdot S^2 + B
\label{parabola}
\end{equation}  
In a system with only Gaussian fluctuations, the above relation is trivially true, reducing to the fixed point $(S = 0, K = 3)$. In a turbulent environment where fluctuacting quantities obey non-Gaussian statistics, the moments spread over finite ranges. Instances of the validity 
of this law may be found in plasma physics \cite{labit,magostini}, atmospheric science \cite{maurizi,alberghi,boundary}, oceanography \cite{sura}, 
laboratory fluids experiments \cite{sreenivasan}. In some cases, there is evidence that a small linear term should be added to Eq. (1) \cite{bertelrud}. 

\begin{figure}
\includegraphics[width=65mm]{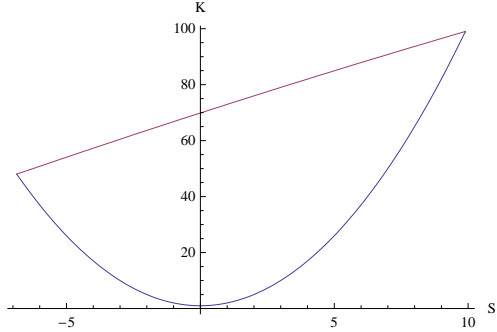}
\caption{The area enclosed between the two curves is the region allowed to all PDFs whose support lies in the interval [-7,10] in units 
of the variance.}
\label{klu}	
\end{figure}

One might wonder about the reasons that lead to the validity of (\ref{parabola}). 
This task has been addressed in connection with specific problems \cite{maurizi,sura}. 
The fact that (\ref{parabola}) is encountered in so many different
situations leads to conjecturing that it must not be model-dependent, rather arises 
because of general constraints, satisfied by several (although not necessarily all) systems encountered in nature. 
Krommes developed an earlier model by Sura and Sardeshmukh \cite{sura}, originally conceived only to 
deal with fluctuations of the sea surface temperature, and showed that it worked equally well the plasma density fluctuations reported in \cite{labit}:
that is, two completely different problems may be modelled starting from the same formalism, hint
of some common underlying physics. 
We believe, however, that there is an even more general rationale 
and in this work are going to propose reasons for this belief.
Several scattered considerations, both of physical as well as purely mathematical nature, 
will be assembled together. Their collection provides a consistent body of evidence in favour of our thesis,
although-of course-a full-fledged proof cannot be obtained.  \\   
Let us start with the simplest example of measurement, where just one scalar variable is measured: $x$, a stochastic variable with its own PDF $P$. 
In order to account for variability in the computed moments, we postulate that $P$ 
be a function, besides of $x$, of some parameters $a_i$: $P = P(x; \{a_i\})$, hence
$<x^m> = \int P x^m dx $ are function of ${a_i}$. The role of $a_i$ is 
modelling the interaction between the system and its environment.  
Let us consider first the case when we have one parameter available: $S = S(a), K = K(a)$
(The zero-parameter case is trivial, since no variability of $S$ and $K$ is then allowed). 
We make the rather natural postulate that the dependence from $a$ is smooth. 
We may suppose that there exists a neighbourhood around $S = 0$ where the relation $S(a)$ is reversible: $ a = a(S)$.
Hence, $K = K(a(S)) \to K(S)$. Because of the postulated smooth dependence from $a$, we may Taylor expand
$K$ around $S = 0$: $ K = K_0 + K^{'} S + ( K^{''}/2) S^2 + ...$. 
Many systems are invariant with respect
to the sign inversion of $x: x \to -x$. For instance, $x$ may stand for the measurement of a 
velocity, as is often the case in fluids dynamics or geosciences. 
An inversion of sign corresponds to the arbitrary choice of the direction of motion. However, $S$ is odd with respect to this operation, while $K$ is even.
Therefore, all coefficients in front of odd powers of $S$ must be null in the above expansion. This yields
\begin{equation}
K = K_0 + ( K^{''}/2) S^2
\label{unop}
\end{equation}
neglecting terms of order $S^4$ or higher. The presence of a linear contribution in (\ref{unop})
could be related to the breaking of the symmetry $ x \to -x $:
i.e., if some constraints do exist in the system preventing $ x \to - x$ be an operation physically realizable 
for the system under consideration. For instance, if $x$ stands for the measurement
of a particle density, only positive values make sense. In this case, there does not
appear to exist justification for discarding the linear term (but for the case of very small fluctuations,
as we shall emphasize later). \\
Being a truncated Taylor expansion, Eq. (\ref{unop}) may only be valid as long as higher order terms remain negligible.
Assessing how large is this range $\Omega$ is an issue that, in principle, 
can be resolved only by actual inspection of each specific system. We provide, however, several considerations
supporting the view that, for most systems, $\Omega$ is as large as the region that may be experimentally scanned.\\
It is well known that, basing upon purely mathematical manipulations, one can provide a lower bound to 
$K$, whatever the PDF \cite{pearson}: 
\begin{equation}
K  \ge S^2 + 1 
\label{kl}
\end{equation}
It is less known that, for PDFs with bounded support, an \textit{upper bound} for $K$ is computable, too \cite{simpson}. This case is highly relevant to our considerations since
fluctuations of infinite amplitude are not physically realizable, hence ultimately all experimental PDFs have finite
support. Let $[l,u]$ be this support, where $l,u$ are measured in units of the standard deviation of the distribution, and $<x> \, = 0$. The two conditions hold \cite{simpson}:
\begin{equation}
K  \le S^2 + 1 - { l\; u \;(S - l + l^{-1}) \;(S - u + u^{-1}) \over 1 + l\; u}
\label{ku} 
\end{equation}
\begin{equation}
l - l^{-1} \le S \le u - u^{-1}
\label{limitis}
\end{equation}  
Eq. (\ref{limitis}) assigns upper and lower bounds achievable by $S$. We may set
$|l|, u > 1$ without much loss of generality (it must be $ u - l > 1$ by construction). Hence, (\ref{limitis}) 
reduces to $ l < S < u$. 
On the other hand, $ l, u $, cannot be exceedingly large, since this would imply that most of the data lie 
in a very narrow interval of values (remember that they are normalized to the variance): 
this is not the signature of a turbulent system. \\
An example of the region allowed to be spanned by a system in the plane $(S,K)$
is shown in the figure (\ref{klu}). 

For systems endowed with mirror symmetry (hence, $-l = u$), we may write the formal expression: 
$K  = G(S^2) S^2 + C(S^2)$ and, as $S \to l,u$, because of (\ref{ku}), $G(S^2)\to 1, C(S^2) \to 1$. 
Empirically, $G(S=0), C(S=0) $ turn usually out to be quite close to unity, too:
$G(0) = 3/2$ \cite{labit,sura}, and $ G(0) = 1.66 $ for the data in fig. \cite{magostini}; the estimate of $C(0)$ is less
precise due to the large vertical scatter of points, but is always of order unity.  
This conforts us in guessing that $G, C$ remain almost constant throughout all the range $(l,u)$ and hence
the parabolic relation between $K$ and $S$ must hold in the same interval. \\
As a second supporting piece of evidence, we note that disparate problems may be collapsed quite often to within a few classes of systems, amenable to analytical treatment. Accordingly, their empirical PDFs can be well approximated by few classes of analytical functions. For example: lognormal PDF, Gamma PDF, etc..., arise in 
all those problems that involve addition of random variables $X$ regardless of the nature of $X$ (say, queuing models, the flow of items through manufacturing and distribution processes, the load on web servers); Poisson 
distribution is related to the waiting times between independent events, and so on. 
In the case of these analytical PDFs, the parameters $a_i$ are simply the free parameters entering the definition of the PDF.  
We restrict to those analytical PDFs where $S,K$, depend upon one single parameter. A scrutiny among these classes of analytical functions shows that Eq. (\ref{unop}) turns out to be an excellent approximation over at least a large $S$ interval, and it is even an \textit{exact} relation, valid for all $S$. We mention, e.g., the problems that lead to Gamma, Inverse Gaussian, Poisson, $\chi^2$ and the Generalized Extreme Value (GEV) Distribution. 
It is straightforward to check that the quadratic relation between $S$ and $K$ is \textit{exact} when the dependence from the parameter $a$ takes a fairly simple expression: $S$ and $K$ may be parameterized as $S \propto a$, $ K  \propto a^2$ plus constant terms. Therefore, although we reached (\ref{unop}) through
a Taylor expansion, its validity is based on more general considerations:
all those sytems for which \textit{(I)} the interaction with the environment may be modelled by means of just one
effective continuously varying parameter $a$, and \textit{(II)} $a$ can be defined such that 
the ``response'' of the system, quantified in terms of departure of $S$ from its Gaussian value, 
is linear with respect to $a$, are expected to obey Eq. (\ref{unop}). \\
Let us move to the case where more parameters drive $S$ and $K$. We will
study the two-parameter case. The other cases are not necessarily a trivial generalization of this one,
but we believe that, even limiting to two parameters, we are able to include a large range
of realistic systems. Let $S = S(a_1, a_2), K = K(a_1, a_2)$.
No such inversion as in the one-parameter case is possible now. However, we can still reduce to that case when the system is ``quasi one-dimensional'', i.e., when the dependence from one of the parameters (say, $a_1$) is much faintier that from the other ($a_2$). Hence, expressing $a_2 = a_2(S,a_1), K = K(a_1,a_2(S,a_1))$
leads to an expression formally identical to (\ref{unop}): $K \approx K_0(a_1) + K^{'}(a_1) S + (1/2) K^{''}(a_1) S^2 + ...$. Therefore, for each fixed value of $a_1$, the curve $K(S)$ is approximately parabolic. Varying $a_1$, we plot on the plane $(S,K)$ different parabolas. If the dependence from $a_1$ is strong, a scan over 
its admissible range of values will lead to plotting parabolas that span all or most of the plane. 
Conversely, if $K_0, K^{'}, K^{''}$ do depend only weakly upon $a_1$, we recover a fan of paraboles close to each other, practically spanning a restricted region of $(S,K)$ plane. 
Therefore, the presence of a weak dependence from a second parameter may account for the spread of the points 
around the fitting parabole that is commonly observed in experiments (see, e.g., \cite{labit}). Actually, this spreading cannot 
be attributed to ``experimental errors'' or other sources of noise: the statistical error due to the finiteness
of the sample can be computed and is negligible in our cases. It is important therefore that our theory be able to explicitly take the spread into account.\\ 
We substantiate the above statements with a few examples where our conjectures may be verified explicitly.
The first example involves the Hasegawa-Mima-Charney (HMC) Equation, that in its non-dissipative version takes the form
\begin{equation}
\left(1 - \nabla^2 \right) {\partial \phi \over \partial t} + U {\partial \phi \over \partial y} - [\phi, \nabla^2 \phi] = 0
\label{HMC}
\end{equation}
where $[...]$ stands for the Poisson bracket. HMC is a two-dimensional nonlinear equation widely known and studied
for its capability of model wave behaviour of such different systems as electrostatic drift waves in magnetized plasmas
and the incompressible motion of shallow rotating neutral fluids (See \cite{horton}, ch. 6): $\phi$ is the electrostatic potential in the first case and the fluctuation of the fluid depth in the second one. It is therefore a fairly good workhorse for a statistical theory of turbulence. 

Horton and Ichikawa made a torough analysis of this equation. Among its solutions (although not exhausting
their whole range) they found that two scenarios may co-exist.
At low amplitudes of the field $\phi$, fluctuations do resolve into a sea of linear non-interacting waves. The field $\phi$ is therefore (even wildly) fluctuating but no longer technically 
turbulent. The Central Limit Theory applies in this case: the measured signal is given by the sum 
of a large number of independent contributions and therefore has Gaussian statistical 
properties.  
The nonlinear high-amplitude part of the fluctuations develops into solitary coherent structures (vortices). 
In order to describe realistic systems interacting with an external environment, the dynamics 
must be augmented with source and dissipative terms, whose relative equilibrium will determine
the average amplitude of fluctuations. Ultimately, therefore, we expect a whole continuum
of its statistical properties, at one extremum including the Gaussian limit.
The peculiar form of these solutions of HMC Equation allows for an explicit computation of its statistical moments (\cite{horton}, par. 6.7). It is convenient to introduce the parameters 
$f_p$ = packing fraction, that is, the fraction of surface occupied by nonlinear coherent structures;
and $A$ = amplitude of the nonlinear part of the fluctuations, that we suppose constant in order to grasp 
simpler results, and normalize with respect
to linear fluctuations: $<\varphi^2> \equiv 1$.  
Finally, 
\begin{equation}
S \approx {-3 f_p A + f_p A^3 \over  (f_p A^2 + 1)^{3/2}} \; , \; 
K \approx {f_p A^4 + 6 f_p A^2 + 3  \over (f_p A^2 + 1)^2} 
\label{eq:hmcks}
\end{equation}  
For fixed $A$ and small $f_p$ (say, $< 0.1$), Eq. (\ref{eq:hmcks}) yields a linear dependence
between $S$ and $K$. The possibility of a linear dependence had to be envisaged because of the lack of symmetry $ A \to -A, f_p \to - f_p$. Conversely, the trend is almost quadratic with $A$ for fixed $f_p$ and moderately 
large $A (< 5)$. It is interesting that,  according to real data \cite{spolaore} $f_p$ is
actually small (  $ < 0.1 \div 0.2 $).  
In realistic situations, several fields are coupled. For instance, in plasma physics HM Equation goes
into two-equations Hasegawa-Wakatami model when non-adiabatic small density fluctuations are included, too.
Increasing the number of fields increases the number of control parameters, too, but each field $F$
depends strongly only upon a subset $a_F$ of all parameters, the remaining ones playing a weaker role, 
hence, qualitatively things are not different from the ``quasi-one-dimensional'' case studied earlier. \\ 
The model proposed by Sura and Sardeshmukh (SS) \cite{sura} has played a main role in our considerations,
hence it is interesting to see how it fits into our picture. We refer the reader to the original paper for the details
and provide here just the fundamental results. SS model reduces basically to just one equation for the time evolution of sea surface temperature fluctuations $\delta T$ in the presence of external fluctuating forcing (heat flux). Part of the forcing ($R$) is due to background unknown sources, and modelled as an additive noise. Another part ($F$) arises as a consequence of the coupling between the sea surface and the atmosphere. Since feedback from the sea over the atmosphere cannot be neglected, $F$ is function of $\delta T$. The random character of the drive converts the evolution equation into a stochastic differential equation with additive and multiplicative noise. The stationary PDF of the temperature fluctuations and all moments are retrieved by solving the associated Fokker-Planck Equation, which can be done analytically. Four parameters are needed: $\sigma_F, \sigma_R$, are the amplitudes of the stochastic forcings; $1/\lambda, 1/\phi_F$, are characteristic times that quantify the rate of relaxation of $\delta T$, $F$ towards steady states (In \cite{sura}, it is used the parameter $\phi$:  
$\phi_F = \phi^2 \times \sigma_F^2$), but two are used by assigning the mean value and the variance. Hence, we remain with two parameters and we expect Eq. (\ref{parabola}) to hold and $A, B$ to be function of one further 
control parameter $a_0$. The explicit calculations of SS yield the
\textit{exact} result, confirming our expectations:
\begin{equation}
K = {3 \over 2} \left( 1 - {a_0 \over 2} \right) S^2 + 3 \left( 1 - a_0 \right),  \quad
a_0 = { \phi_F \over 2 \phi_F -\lambda}  
\end{equation}

Summarizing, we claim that the parabolic relation (\ref{parabola}) arises because of the  
validity of the following conditions: \textit{(A)}  fluctuating  systems may ultimately be modelled by stochastic differential equations (SDE); \textit{(B)} the interaction with the external environment is phenomenologically
fed into the SDE through a number of effective parameters that, ultimately, enter into the 
definition of the moments $S, K$. It is often possible to identify a single parameter $S,K$ depend strongly
upon, and a small number of secondary parameters. This does not appear an exceedingly demanding requisite. On the contrary, most if not all systems are modelled through equations that depend on a very small
number of parameters. The paradigm is the Navier-Stokes equation that, once in dimensionless form,
admits as external parameter the Reynolds $Re$ number. 
We were not able to find and explicit study of
skewness versus kurtosis for fluid turbulence driven by Navier-Stokes equation, but several
researchers addressed the issue of the scalings $S, K $ \textit{versus} $Re$. A review with data is \cite{sreenivasan}.
Figs. (5,6) of that paper show that both $S$ and $K$ scale with $Re$ for rather large values
of this parameter ($Re > 100$): $ S, K \propto Re^{a_S , a_K}$.
Hence, $K \propto S^{(a_K/a_S)}$, and $a_K/a_S$ is rather close to 2 (the best fit being 2.5). 
\textit{(C)} Under a very weak external drive,   many (although definitely not all)
systems collapse to their nonturbulent Gaussian limit, a linear superposition of independent oscillations, and correspondingly 
$S = 0, K = 3$. Finite driving displaces
smoothly the system from this condition, and the resulting  signal  is a combination of Gaussian and nonlinear non-Gaussian fluctuations. The driving itself must be implemented into a parameter, $a$. 
A linear response ansatz makes $S$ proportional to $a$. This, together
with the next point (D), leads to the \textit{exact} validity of (\ref{unop}). 
\textit{(D)} The linear term in (\ref{parabola}) is absent because the system
studied has intrinsic mirror symmetry $ x \to - x$.  Another possibility is that 
only small fluctuations $\delta x$ around an equilibrium
state are investigated. In this latter case, the full system may not possess mirror symmetry, but 
the reduced one does: $\delta x \to - \delta x$. In this case,
one may be led to condition (C) if the zero-fluctuation limit of the system is not turbulent and the small fluctuations slightly depart the system from Gaussian statistics, making Eq. (\ref{unop}) sensible.  
\textit{(E)} Finally, purely mathematical constraints exist, arising just out of the definition 
of $S, K$, and the fact that physically realizable systems are always finite, that prevent in any case
this couple of parameter to depart sensitively from the scaling (\ref{parabola}).  

In summary, therefore, our claim is that the parabolic relation between $S$ and $K$ encountered in the statistical
treatment of data from turbulent systems is not likely to provide relevant informations about the underlying physics. 

\begin{acknowledgments}
This work was supported by the European Communities under
the contract of Association between EURATOM/ENEA. 
D. Escande and E. Martines read several versions of the draft and suggested invaluable corrections and improvements.
\end{acknowledgments}

\end{document}